\documentclass[CEJP,DVI]{cej} 
\usepackage{layout}
\usepackage{amsmath}
\usepackage{textcomp}
\usepackage{hyperref}

\title{Measurements of Heavy Flavor and Di-electron Production at STAR}

\articletype{Editorial}

\author{Gang Wang (STAR Collaboration)\inst{1}\email{gwang@physics.ucla.edu}}

\institute{
     \inst{1} Department of Physics and Astronomy, University of California, Los Angeles, USA }

\abstract{Heavy quarks are produced early in the relativistic heavy ion collisions,
and provide an excellent probe into the hot and dense nuclear matter created at RHIC.
In these proceedings, we will discuss recent STAR measurements of heavy flavor production,
to investigate the heavy quark interaction with the medium. 
Electromagnetic probes, such as electrons, provide information on the various stages
of the medium evolution without modification by final stage interactions.
Di-electron production measurements by STAR will also be discussed.
}


\begin{document}
\maketitle


\section{Heavy Flavor}

Heavy quarks, charm and bottom, are believed to be produced mostly via initial gluon fusion~\cite{fusion}
in nuclear collisions at the Relativistic Heavy Ion Collider (RHIC), 
and their propagation through the medium created by the collisions is an excellent probe to study the medium properties.
Models that describe the experimental results of heavy-ion collisions often assume that light quarks in the medium reach
thermalization on an extremely short timescale ($\sim 0.5$ fm/$c$)~\cite{time}.
However, the reason for the fast thermalization is still elusive, and the extent of the thermalization is also unclear.
With much larger masses than light quarks, heavy quarks are expected to thermalize much more slowly than light partons.
Thus by measuring the collective motion ($v_2$) of heavy flavor particles like $J/\psi$ and $D$ mesons,
we can study the flow of charm quarks and the extent of thermalization reached by the medium.

\begin{figure}
\begin{minipage}[c]{0.48\textwidth}
\center
\includegraphics[width=\textwidth]{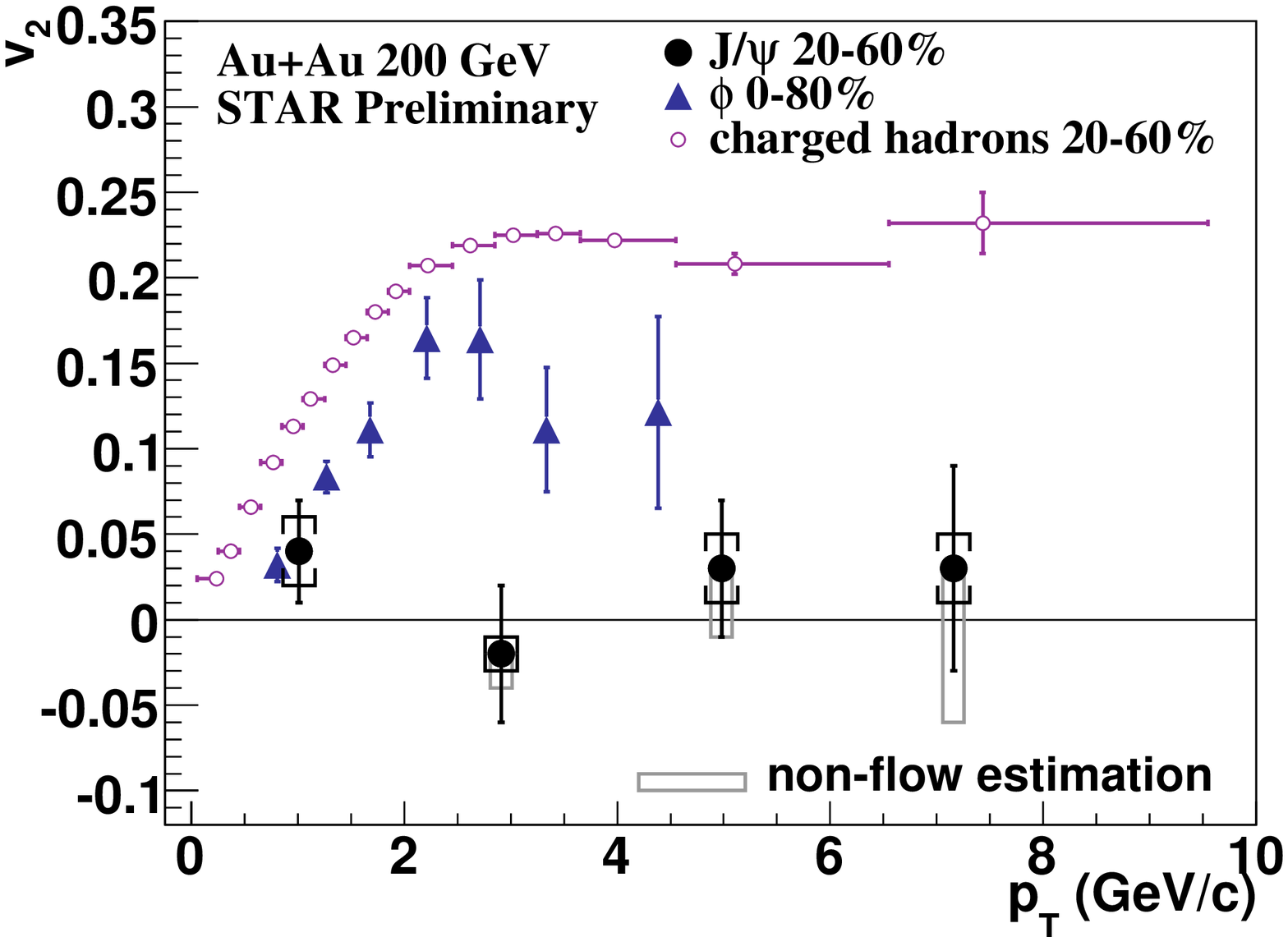}
\vspace{-0.06cm}
  \caption{$J/\psi$ $v_2$ as a function of $p_T$ for $20-60\%$ Au+Au collisions at 200 GeV. $v_2$ results of $\phi$ and charged hadrons are also shown for comparison~\cite{zebo}.
}
\label{fig:1}
\end{minipage}
\begin{minipage}[c]{0.48\textwidth}
\center
\vspace{-0.5cm}
\includegraphics[width=\textwidth]{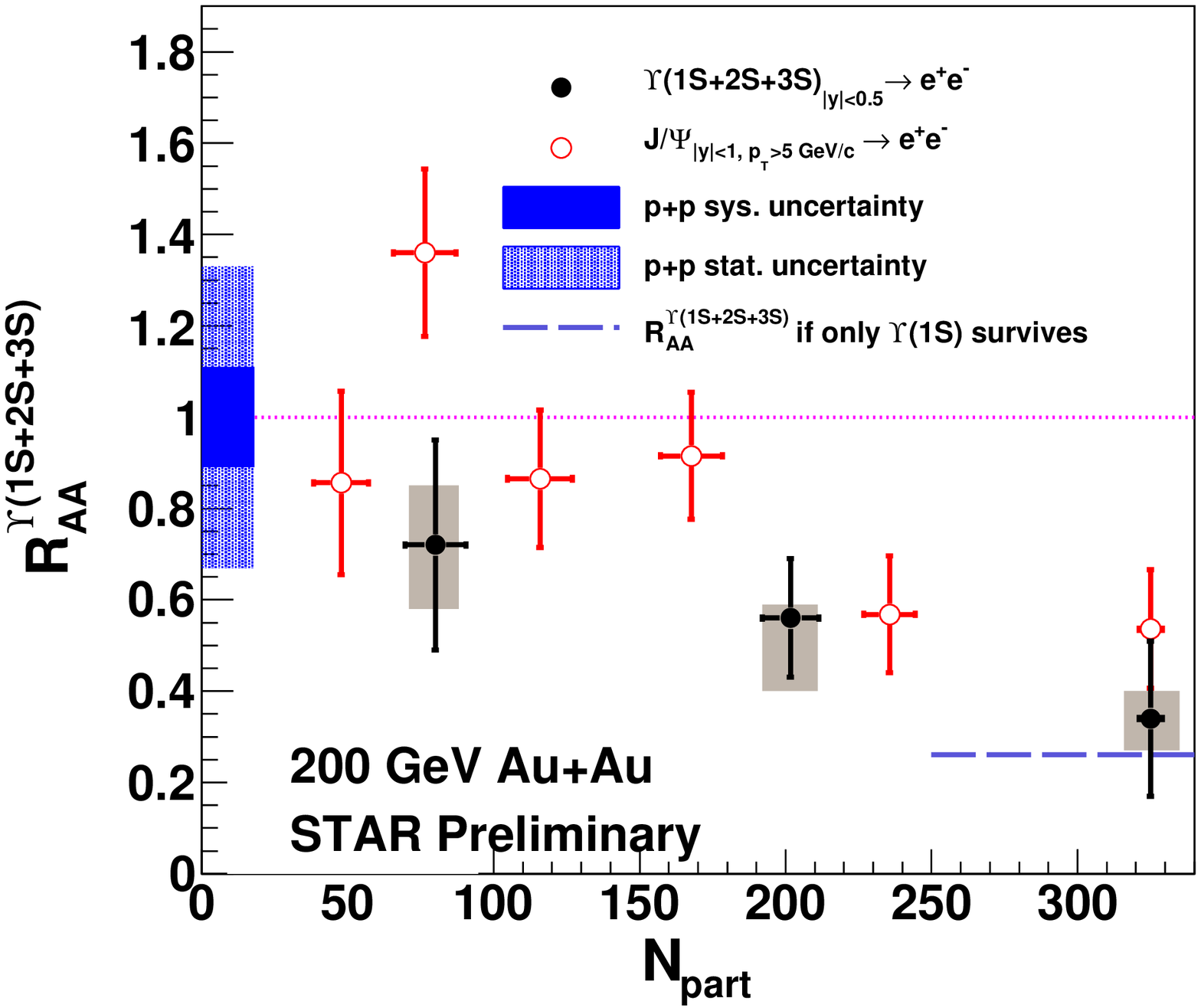}
  \caption{$R_{AA}$ of $\Upsilon$(1S+2S+3S) and high $p_T$ $J/\psi$ as a function of centrality for 200 GeV Au+Au collisions~\cite{zebo,rosi}.
        }
\label{fig:2}
\end{minipage}
\end{figure}

Figure~\ref{fig:1} shows STAR measurements of $J/\psi$ $v_2$ as a function of transverse momentum ($p_T$) for $20-60\%$ Au+Au 
collisions at 200 GeV~\cite{zebo}.
$J/\psi$ was reconstructed from the $e^+e^-$ channel.
At low $p_T$, $J/\psi$ $v_2$ could be positively finite, the confirmation of which requires further measurements 
with higher precision.
For $p_T > 2$ GeV/$c$, unlike $v_2$ of charged hadrons or $\phi$ mesons, $J/\psi$ $v_2$ is consistent with zero,
disfavoring the picture that high $p_T$ $J/\psi$ is produced dominantly by coalescence from thermalized charm 
and anti-charm quarks.
To ascertain whether coalescence fails due to the low production cross section of charm quarks or
charm quarks do not flow due to their little interaction with the medium, we need to measure
$v_2$ of open heavy flavor. STAR is carrying out the detector upgrade of the Heavy Flavor Tracker (HFT),
which will help answer the question on charm flow in the near future~\cite{Jan}.

The nuclear modification factor ($R_{AA}$) is a potential measure for the quarkonium suppression, and their possible
interaction with the medium.
$R_{AA}$ of high $p_T$ $J/\psi$ is shown in Fig.~\ref{fig:2} as a function of centrality for 200 GeV Au+Au collisions~\cite{zebo}.
In peripheral and mid-peripheral collsions, $R_{AA}$ is consistent with one, seemingly to display no $J/\psi$
suppression at high $p_T$. However, the $J/\psi$ production is influenced by the interplay
of hot and cold nuclear matter effects, which complicates our understanding.
In central collisions, $R_{AA}$ is around $0.6$. But this does not guarantee a significant $J/\psi$-medium interaction,
because there could be about $40\%$ of $J/\psi$ coming from feeddown of excited states and $B$ decay in p+p collisions,
while excited states and $B$ mesons could be melted in the medium leaving prompt $J/\psi$ intact in Au+Au collisions.
The HFT is needed to single out the prompt $J/\psi$ from displaced feeddown $J/\psi$.
A similar case is $\Upsilon$(1S+2S+3S) $R_{AA}$~\cite{rosi}, also illustrated in Fig.~\ref{fig:2}.
The $\Upsilon$(1S+2S+3S) is significantly lower than one in central collisions,
and within errors close to the estimate where only $\Upsilon$(1S) survives the medium temperature.
STAR is building a new detector subsystem, the Muon Telescope Detector (MTD)~\cite{lijuan}, to reconstruct quarkonia from
the di-muon channel. The better mass resolution from di-muon, together with higher luminosity,
 will allow us to separate different $\Upsilon$ states.

\begin{figure}
\begin{minipage}[c]{0.48\textwidth}
\center
\vspace{0.15cm}
\includegraphics[width=\textwidth]{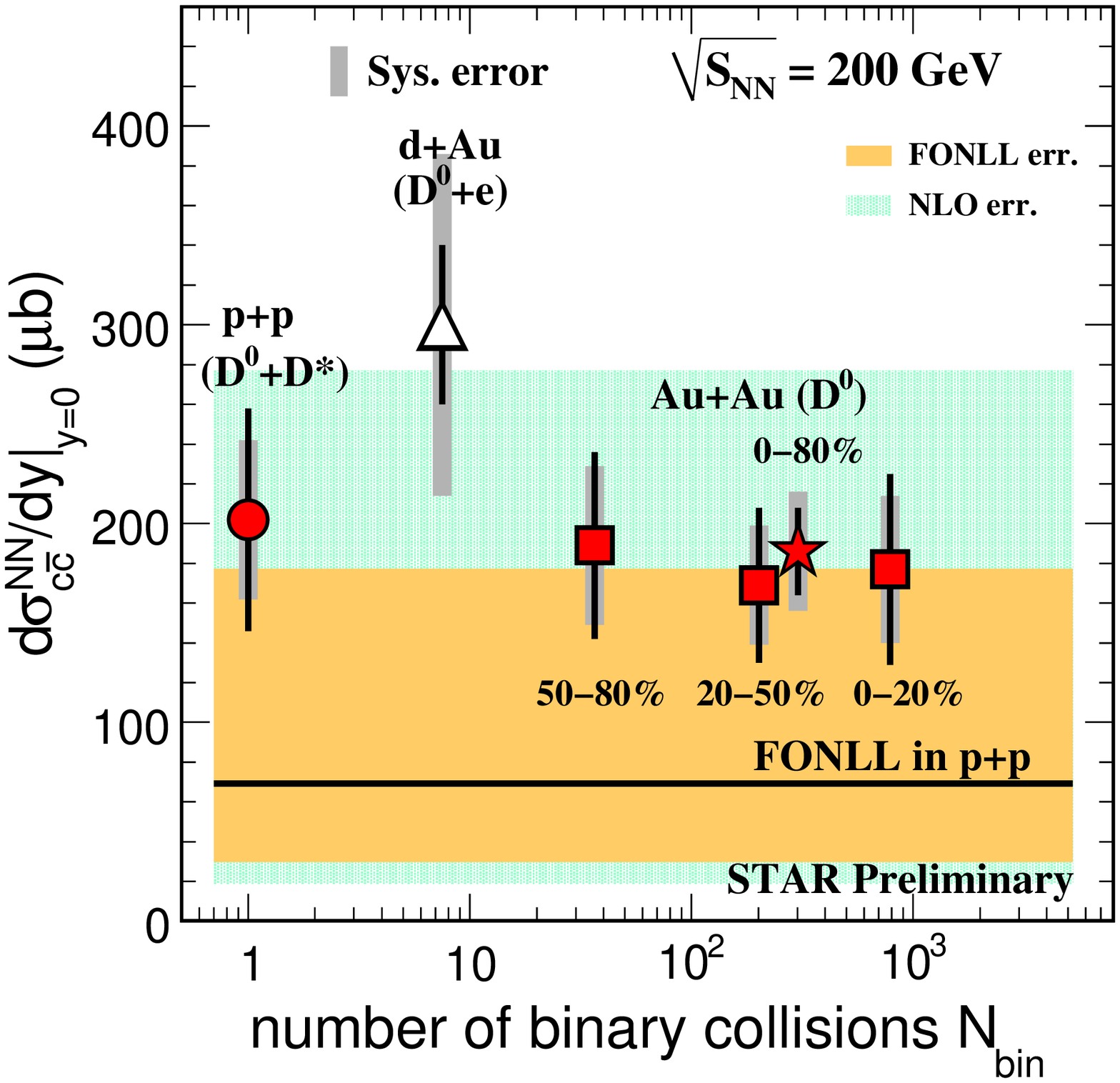}
  \caption{The charm production cross section per nucleon-nucleon collision at mid-rapidity as a function of $N_{bin}$~\cite{yifei,yifei2}.
}
\label{fig:3}
\end{minipage}
\begin{minipage}[c]{0.48\textwidth}
\center
\includegraphics[width=\textwidth]{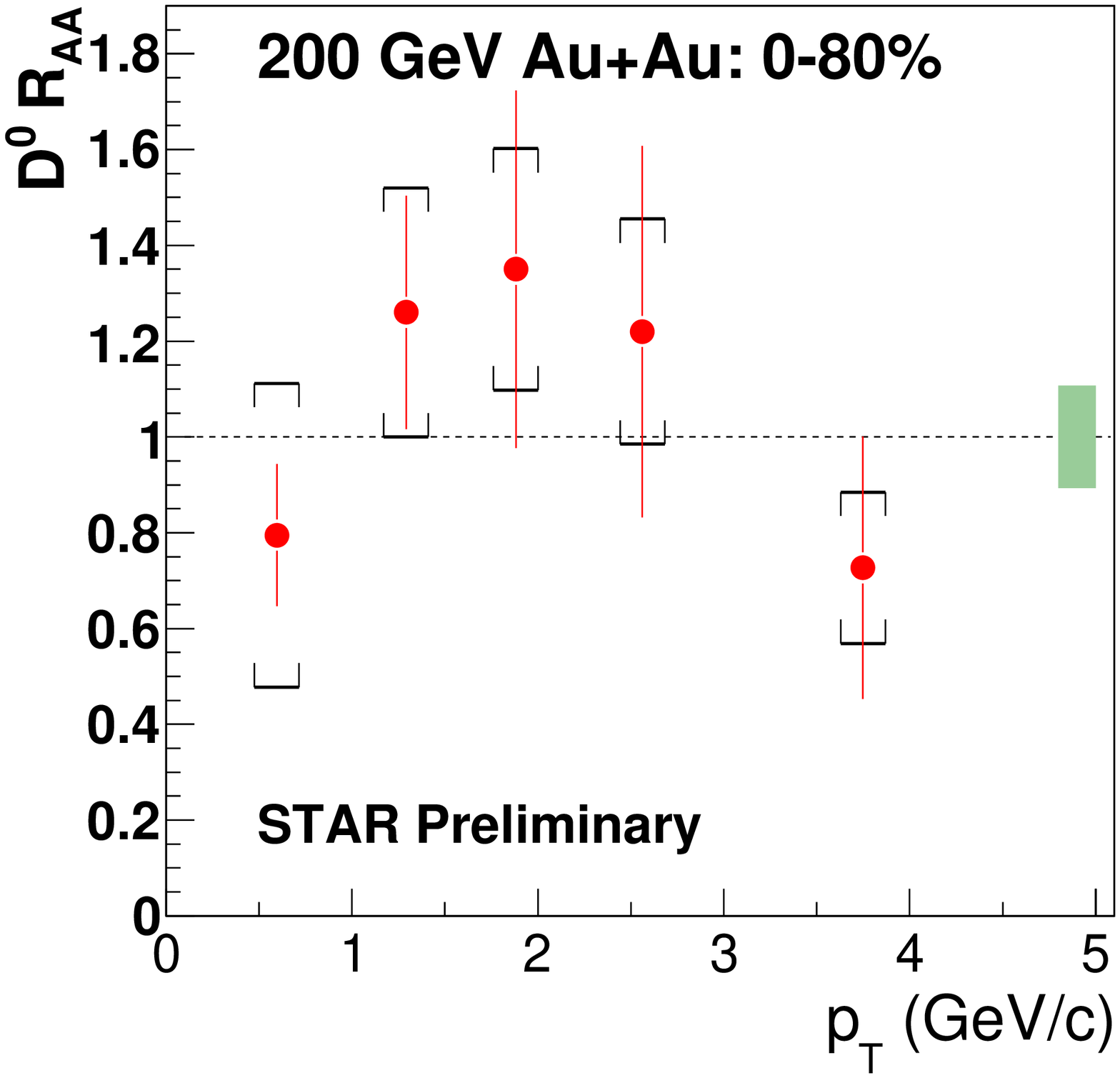}
  \caption{$D^0$ $R_{AA}$ as a function of $p_T$ for $0-80\%$ Au+Au collisions at 200 GeV~\cite{yifei}.
        }
\label{fig:4}
\end{minipage}
\end{figure}

More than $99\%$ of charm quarks hadronize into open charm hadrons, so $D$ meson measurements
are crucial to the determination of the charm production cross section.
Figure~\ref{fig:3} shows STAR measurements of the charm production cross
section per nucleon-nucleon collision at mid-rapidity as a function of number of binary collisions ($N_{bin}$),
for p+p~\cite{yifei}, d+Au~\cite{yifei2} and Au+Au~\cite{yifei} collisions at 200 GeV.
$D$ mesons were reconstructed from the hadronic decay channels.
Within errors, the results are in agreement with each other and follow the $N_{bin}$
scaling, indicating that charm quarks are produced via initial
hard scatterings at an early stage of the collisions at RHIC.

$D^0$ $R_{AA}$~\cite{yifei} was obtained via dividing $D^0$ yields in $0-80\%$ Au+Au collisions by
the power-law fit to p+p yields scaled by $N_{bin}$, shown in Fig.~\ref{fig:4}.
No suppression is observed at $p_T < 3$ GeV/$c$. From previous $R_{AA}$ measurements of non-photonic
electrons (NPE) from $D$ and $B$ decays~\cite{npe}, we may speculate that a significant $D^0$ suppression
should occur at high $p_T$.
Also, from NPE $v_2$ measurements~\cite{NPEv2}, we expect finite $D^0$ $v_2$.
The future STAR HFT provides the necessary resolution to reconstruct secondary vertices of $D$ mesons
and enable high precision measurements of $D^0$ $v_2$ and high $p_T$ $D^0$ $R_{AA}$,
to address the light flavor thermalization and charm quark energy loss mechanisms.
The more accurate charm measurement also leads to the more direct determination
of the relative contributions of charm and bottom quarks to NPE, previously explored via
NPE-hadron correlations~\cite{NPE-h}. The bottom contribution to NPE, in turn,
will help determine the bottom production cross section.

\begin{figure}
\begin{minipage}[c]{0.48\textwidth}
\center
\vspace{-0.16cm}
\includegraphics[width=\textwidth]{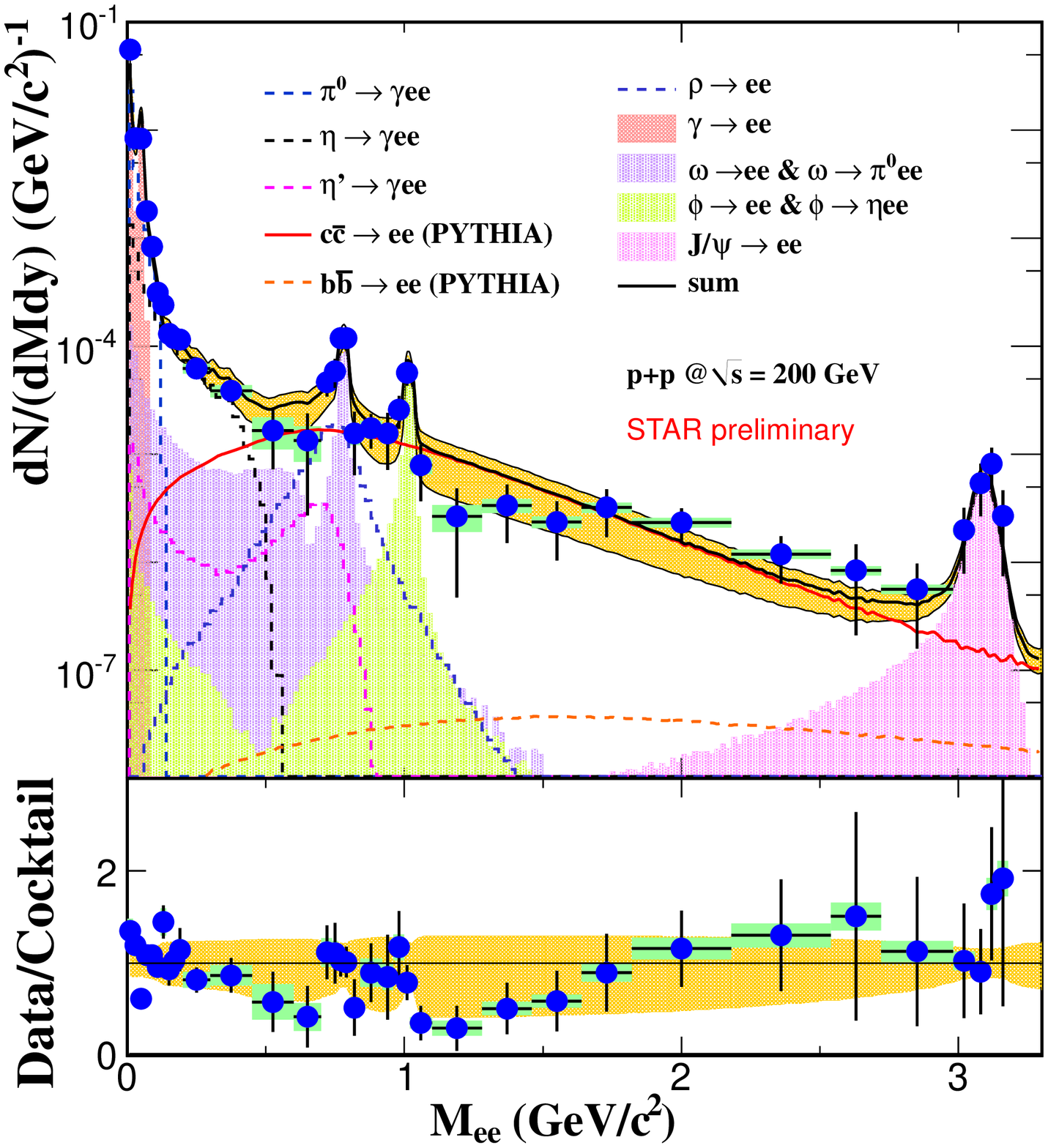}
  \caption{Invariant mass spectra from 200 GeV p + p collisions~\cite{jie}. 
	   The yellow band is systematic error on cocktail, and green boxes are systematic error on data.
}
\label{fig:5}
\end{minipage}
\begin{minipage}[c]{0.48\textwidth}
\includegraphics[width=\textwidth]{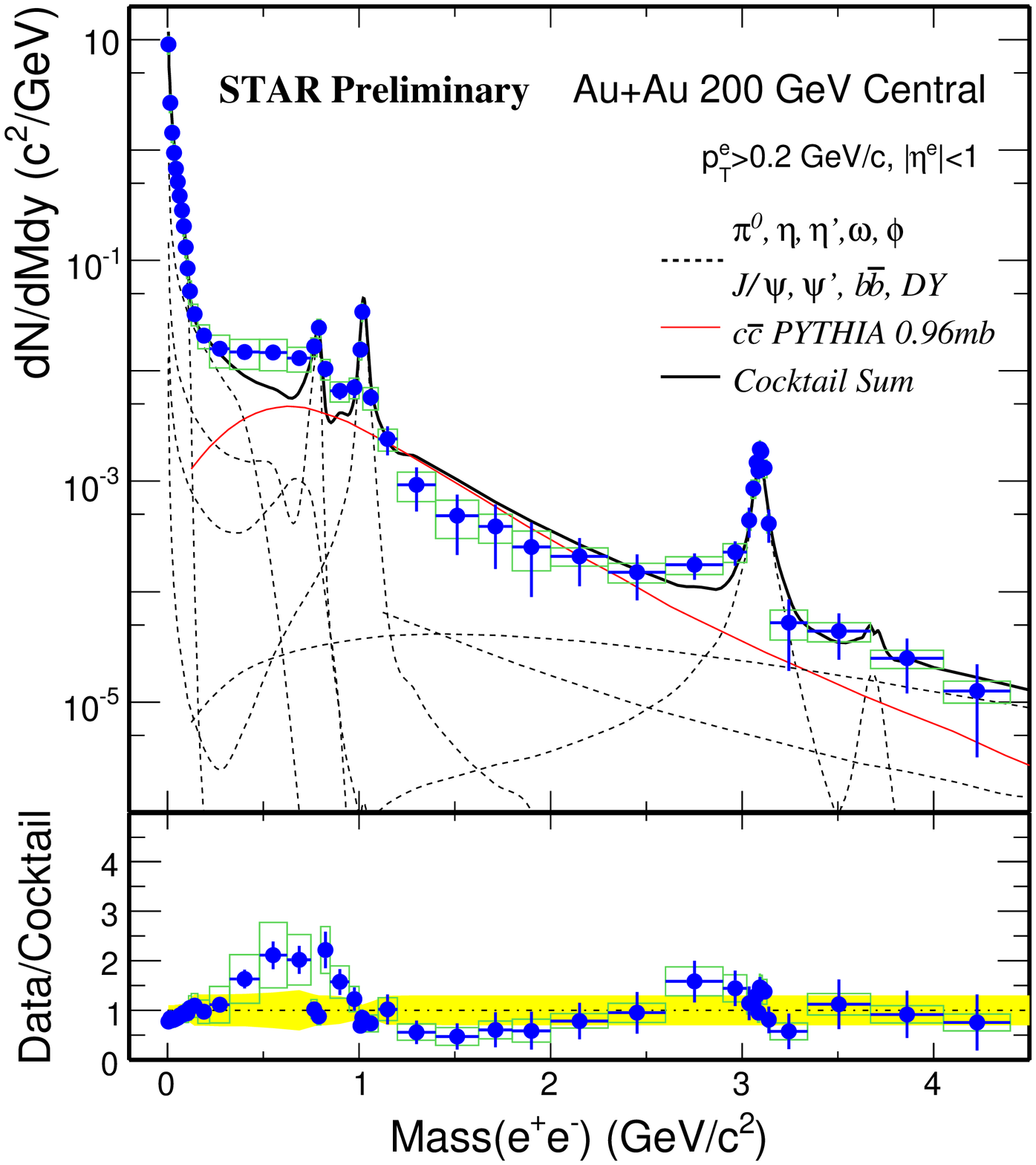}
  \caption{Invariant mass spectra from central Au+Au central collisions at 200 GeV~\cite{jie}. 
	   The yellow band is systematic error on cocktail, and green box is systematic error on data.
        }
\label{fig:6}
\end{minipage}
\end{figure}

\section{Di-electron}
Di-electron distributions provide a penetrating probe for the entire evolution of the hot
and dense nuclear medium, and the di-electron sources vary as a function of kinematics.
In the low mass region (LMR: $m_{ee} < 1.1$ GeV/$c^2$), direct photons and in-medium properties of vector mesons
can be studied through their di-electron decays, 
while in the intermediate mass region (IMR: $1.1 < m_{ee} < 3$ GeV/$c^2$) quark gluon plasma (QGP) radiation 
is expected to have a significant contribution at RHIC~\cite{IMR}. 
In the high mass region (HMR: $m_{ee} > 3$ GeV/$c^2$), di-electrons are mostly from heavy quark decays and Drell-Yan processes.

With the completion of the full barrel Time-Of-Flight (TOF) detector~\cite{tof}, the electron
identification has been significantly improved at STAR, especially in low momentum region.
The background reconstruction in this analysis is based on the mixed-event
technique (for IMR) and the like-sign method (for LMR). 
The di-electon continuum results are within the STAR acceptance ($p^e_T > 0.2$ GeV/$c$ , $|\eta^e| < 1.0$ and 
$|y^{ee}| < 1.0$ ) and corrected for efficiency.

The di-electron invariant mass spectra from 200 GeV p+p and central Au+Au collisions
are shown in Fig.~\ref{fig:5} and Fig.~\ref{fig:6}, respectively~\cite{jie}.
The results from p + p collisions are consistent with hadron decay cocktail
simulation~\cite{cocktail}, providing a baseline for Au+Au collisions.
For central Au+Au collisions, there is an enhancement with a factor of
$1.72 \pm 0.10^{stat} \pm 0.50^{sys}$ compared with cocktail (without $\rho$) in LMR.
In the lower panel of Fig.~\ref{fig:6}, the bump, if any, around the mass of $\rho$ meson,
is much broader than that caused by $\rho$ mesons in the vacuum,
possibly indicating the medium modification of the $\rho$ mass peak.
Future measurements with increased statistics will improve our understanding on the LMR vector meson production.
In the IMR, the yield from binary-scaled charm contributions from PYTHIA seems to 
be systematically higher than the data in central collisions with uncertainties touching each other.
However, these predictions do not take into account the modification of charm production in the medium.
The modification would soften the predicted mass spectra, and leave more room
for the thermal production in our data. Note that improvements
of both data and theoretical calculations are needed to have a conclusive comparison.
The future HFT and MTD upgrades will help us to separate the charm and thermal radiation contributions,
for example via the $e-\mu$ correlation~\cite{lijuan2}.

\section{Outlook}
Besides the detector upgrades, STAR also carried out the heavy flavor and di-electron measurements
at lower beam energies, for example in 39 GeV Au+Au collisions~\cite{39GeV1,39GeV2}.
The RHIC Beam Energy Scan (BES)~\cite{BES} program aims to search for the onset of deconfinement and
the possible critical point in the QCD phase diagram, and the heavy flavor and di-electron
production measurements could reflect the beam-energy dependency of the medium properties.


\begin{thebibliography}{99}
\bibitem{fusion} J.~Rafelski and B.~Muller, Phys. Rev. Lett. 48, 1066 (1982)
\bibitem{time}  P.~Kolb and U.~Heinz, nucl-th/0305084; P.~Huovinen and P.~V.~Ruuskanen,
Annu. Rev. Nucl. Part. Sci. 56, 163 (2006)
\bibitem{zebo} Z.~Tang et al. (STAR Collaboration), J. Phys. G 38, 124107 (2011)
\bibitem{Jan} J.~Kapitan et al. (STAR Collaboration), Eur. Phys. J. C 62, 217 (2009)
\bibitem{rosi} R.~Reed et al. (STAR Collaboration), J. Phys. G 38, 124185 (2011)
\bibitem{lijuan} L.~Ruan et al. (STAR Collaboration), arXiv:0805.4638 
\bibitem{yifei} Y.~Zhang et al. (STAR Collaboration), J. Phys. G 38, 124142 (2011)
\bibitem{yifei2} J.~Adams et al. Phys. Rev. Lett. 94, 062301 (2005)
\bibitem{npe} B.~I.~Abelev et al. (STAR Collaboration), Phys. Rev. Lett. 106, 159902(E) (2011)
\bibitem{NPEv2} G.~Wang et al. (STAR Collaboration), Nucl. Phys. A 855, 319 (2011) 
\bibitem{NPE-h} M.~M.~Aggarwal et al. (STAR Collaboration), Phys. Rev. Lett. 105, 202301 (2010)
\bibitem{IMR} R.~Rapp, Phys. Rev. C 63, 054907 (2001) and references therein
\bibitem{tof} STAR TOF proposal, http://drupal.star.bnl.gov/STAR/files/future/proposals/tof-5-24-2004.pdf
\bibitem{jie} J.~Zhao et al. (STAR Collaboration), J. Phys. G 38, 124134 (2011)
\bibitem{cocktail} L.~Ruan et al. (STAR Collaboration), Nucl. Phys. A 855, 269 (2011)
\bibitem{lijuan2} L.~Ruan et al. (STAR Collaboration), Journal of Physics G: Nucl. Part. Phys. 36, 095001 (2009)
\bibitem{39GeV1} Z.~Tang, XXII International Conference on Relativistic Nucleus-Nucleus Collisions (2011)
\bibitem{39GeV2} P.~Huck, these proceedings
\bibitem{BES} M.~M.~Aggarwal et al. (STAR Collaboration), arXiv:1007.2613
\end{thebibliography}
\end{document}